\documentstyle[12pt]{article}

\def\bc{\begin{center}}
\def\ec{\end{center}}
\def\be{\begin{equation}}
\def\ee{\end{equation}}
\def\myappendix{\par
 \setcounter{section}{0}
 \setcounter{subsection}{0}
 \setcounter{equation}{0}
 \setcounter{table}{0}
 \def\appendixname{Appendix}
 \def\appesection{\setcounter{equation}{0}\section}
 \def\@thesection{\Alph{section}}
 \def\thesection{\appendixname\hskip 1.10ex\Alph{section}}
 \def\thesubsection{\@thesection.\arabic{subsection}}
 \def\theequation{\@thesection.\arabic{equation}}
 \def\thetable{\@thesection.\arabic{table}}}

\newcommand{\labar}{\overline{\Lambda}}
\newcommand{\lb}{\overline{\Lambda}}

\newcommand{\kkinetic}{\bar{h}(x)\, \vec{D}^{2}\, h(x)}
\newcommand{\kkrinetic}{\bar{h}(x)\, \vec{D}^{2}_{S}\, h(x)}
\newcommand{\energy}{\bar{h}(x)\, D_{4}\, h(x)}
\newcommand{\unity}{\bar{h}(x)\, h(x)}

\newcommand{\beq}{\begin{equation}}
\newcommand{\eeq}{\end{equation}}
\newcommand{\beqn}{\begin{eqnarray}}
\newcommand{\eeqn}{\end{eqnarray}}

\def\vdir{v\kern-7.8pt\Big{/}}
\def\pdir{p\kern-7.8pt\Big{/}}

\newcommand{\deltambar}{\delta\overline{m}}
\newcommand{\err}[2]{\raisebox{0.08em}{\scriptsize {$\;\begin{array}{@{}l@{}}
                          \plus\makebox[0.55em][r]{#1} \\[-0.12em]
                          \minus\makebox[0.55em][r]{#2}
                        \end{array}$}}}
\newcommand{\plus}{\makebox[15pt][c]{$+$}}
\newcommand{\minus}{\makebox[15pt][c]{$-$}}
\begin{document}
\pagestyle{empty}
\vspace{-0.6in}
\begin{flushright}
CERN-TH.7521/94 \\
ROME prep. 94/1071 \\
SHEP 94/95-14
\end{flushright}
\vskip 0.2 cm
\centerline{\LARGE{\bf{First Lattice Calculation of}}}
\vskip 0.2cm
\centerline{\LARGE{\bf{The B-meson Binding and Kinetic Energies.}}}
\vskip 0.2cm
\centerline{\bf{M. Crisafulli$^1$, V. Gim\'enez$^{2}$,
G. Martinelli$^{3,*}$ and C. T. Sachrajda$^4$}}
\centerline{$^1$ Dip. di Fisica, Univ. ``La Sapienza"  and}
\centerline{INFN, Sezione di Roma, P.le A. Moro, I-00185 Rome, Italy.}
\centerline{$^2$ Dep. de Fisica Teorica and IFIC, Univ. de Valencia,}
\centerline{Dr. Moliner 50, E-46100, Burjassot, Valencia, Spain.}
\centerline{$^3$ Theory Division, CERN, 1211 Geneva 23, Switzerland.}
\centerline{$^4$ Dept. of Physics,   University of Southampton,
 Southampton SO17 1BJ,UK.}
\abstract{
We present the first lattice calculation of the B-meson binding energy
$\labar$  and of the kinetic energy $-\lambda_1/2 m_Q$
of the heavy-quark inside
the pseudoscalar  B-meson. This calculation has required
the non-perturbative subtraction of the
power divergences present  in matrix elements
of  the Lagrangian operator  $\bar h D_4 h$ and
of  the kinetic energy operator $\bar h \vec D^2 h$. The non-perturbative
renormalisation of the relevant  operators
has been implemented by imposing suitable renormalisation
 conditions on quark matrix elements, in the Landau gauge.
Our numerical results have been obtained from several  independent numerical
simulations at $\beta=6.0$ and $6.2$, and using,  for
the meson correlators, the results obtained by the
APE group at the same values of $\beta$. Our best estimate, obtained
by combining results at different values of $\beta$, is
$\labar =190 \err{50}{30}$ MeV.
For the $\overline{MS}$ running mass,
we obtain $\overline {m}_b(\overline {m}_b)
=4.17 \pm 0.06$ GeV, in reasonable agreement with
previous determinations.  From a subset of 36 configurations,
we were only able to establish a
loose upper bound on the $b$-quark kinetic energy
in a $B$-meson, $\lambda_1=\langle B
\vert \bar h \vec{D}^{2} h \vert B \rangle /(2 M_B )<$~1\, GeV$^2$.
This shows that a much larger statistical sample is needed to determine this
important parameter.}
\vskip 0.3cm
\centerline{$^*$ On leave of absence  from Dip. di Fisica, Universit\`a
degli Studi ``La Sapienza", Rome, Italy. }
\vfill\eject
\pagestyle{empty}\clearpage
\setcounter{page}{1}
\pagestyle{plain}
\newpage
\pagestyle{plain} \setcounter{page}{1}

\section{Introduction} \label{intro}

The Heavy Quark Effective Theory (HQET)  \cite{voshi}--\cite{Neubert}
has proven to be an extraordinary tool for studying  heavy flavour
physics.  In this approach, physical quantities are expanded as
series in inverse powers of the heavy quark masses. The  spin-flavour
symmetries, appearing  in the infinite mass limit,  are then   used to
relate different hadron masses or weak amplitudes which control heavy
meson and baryon   decays  \cite{Neubert}. For example, in the infinite
mass limit, the set of six hadronic form factors, which parameterize
the  matrix elements of the flavour changing vector and axial vector
current in $B \to D,D^*$ semileptonic decays, can be reduced  to a
single universal one: the so called the Isgur-Wise function
\cite{iswi}.  Spin-flavour symmetries,  however, are not sufficient to
predict all the properties of the weak form factors and of  other
important quantities  such as the meson decay constants and the
velocity
dependence of the Isgur-Wise function. Among the quantities
that cannot be predicted on the basis of the HQET only, there are
several  parameters which characterize the dynamics of strong
interactions, such as the heavy quark binding energy, relevant for
higher order corrections to the semileptonic form factors, and the
heavy quark kinetic energy, which enters in the  predictions of many
inclusive decay rates  \cite{Neubert,lambdabar}.

The lattice formulation of the HQET offers the possibility of a
numerical, non\--per\-tur\-ba\-ti\-ve  determination of these
quantities from first principles and without free
parameters~\cite{fb,slope}.
For example,  the most important achievement of
lattice simulations of the HQET has been the computation of the B-meson
decay constant in the static limit, $f^{stat}_{B}$.  In this work, we
present the first lattice calculation of the B-meson binding energy,
$\labar$, and of the kinetic energy of the heavy quark  in the B-meson
$-\lambda_1/(2 m_Q)$, where $\lambda_1 =  \langle B \vert \bar h
\vec{D}^{2} h \vert B \rangle/(2 M_B)$.

The parameter  $\labar$ denotes the asymptotic  value of the difference
between the hadron and the heavy quark ``pole" mass $m_Q$
\be \labar\, =\, \lim_{m_{Q} \rightarrow \infty}\,
 \left( M_{H}\, -\, m_{Q} \right) .
\label{eq:naivedef}
\ee
It has been recently shown that the pole mass is ambigous due to the
presence of infrared renormalon singularities  \cite{beneke,bigi}. At lowest
order in $1/m_Q$, the infrared renormalon ambiguity
 appearing in the definition of the pole mass is closely related to
the ultra-violet renormalon singularity present in the matrix elements of
the operator $\bar h D_4 h$.
This singularity is due to the linear  power divergence
  of  $\bar h D_4 h$,  induced by  its mixing
with the lower dimensional operator $\bar h h$. In  perturbation
theory, using  dimensional regularization, the power divergence is hidden by
the absence of
an intrinsic scale in the computation. On the lattice, because
of the hard cut-off,
renormalon poles are absent \cite{bigi}. In this case, the linear divergence
manifests itself as a power  divergence in the inverse lattice spacing $1/a$,
which appears in the  mixing coefficient of the operator $\bar h h$. In
ref.  \cite{mms} it was stressed that these divergences must be subtracted
non-perturbatively since factors such as
\be \frac{1}{a}\exp \Bigl( - \int^{g_0(a)} \frac{d
g^\prime}{\beta(g^\prime)} \Bigr) \sim \Lambda_{{\rm QCD}},
\label{eq:argu} \ee
which do not appear in perturbation theory,  give non-vanishing
contributions as $a \to 0$
(see also refs.~\cite{mmrt,tsuka}).
In this sense, power divergences in theories with a hard cut-off and
renormalon poles in dimensional regularization are closely related. For
a more detailed discussion see ref.~\cite{invisible}.  The intrinsic
ambiguity of $O(\Lambda_{{\rm QCD}})$,  present in the  renormalisation
of  $\bar h D_4 h$, implies an ambiguity in the definition  of a finite
$\labar$ and hence of $m_Q$.

Falk, Neubert and Luke  \cite{fnl}  have proposed a  different  definition
of $\labar$,
\be
\labar \, =\,
\frac{-\langle 0|\,\bar h\Gamma D_4\,q\, |M_{H}\rangle}
{\langle 0|\,\bar h\Gamma q\, |M_{H}\rangle}
\label{eq:fnldef}
\ee
where $h$ ($q$) is the effective heavy-quark (light-quark) field, $\Gamma$ is
a Dirac matrix and $M_{H}$ a meson annihilated (created) by the operator
$J_\Gamma=\bar h\Gamma q$ ($J^\dagger_\Gamma=\bar q\Gamma h$).
This definition contains the same renormalon ambiguities as
that in eq. (\ref{eq:naivedef}).

In the lattice HQET, the ``binding energy" computed in numerical
simulations corresponds to the definition given in  eq.
(\ref{eq:fnldef}). Consider the two-point function
\be
 C(t)= \sum_{\vec x}\, \langle 0|\, J_\Gamma(\vec x,t)
J_\Gamma^\dagger(\vec 0,0)\,|0\rangle
  = \sum_{\vec x}\, \langle 0|\,\bar h(\vec x,t)\Gamma
q(\vec x,t)\; \bar q(\vec 0,0) \Gamma h(\vec 0,0)\,|0\rangle
\label{eq:cdef}
\ee
For sufficiently large Euclidean time $t$,
\be
C(t)\rightarrow Z^2 \exp(-{\cal E}t)
\label{eq:ctbig}
\ee
where $Z$ is a constant. The definition (\ref{eq:fnldef}) implies that
$\labar = {\cal E}$: indeed  ${\cal E}$  can be interpreted as the
difference $M_H-m_Q$ where $M_H$ is the mass of the lightest meson
which can be created by the operator $J_\Gamma^\dagger$. It is clear,
however, that ${\cal E}$ cannot be a ``physical"  quantity because it
diverges linearly as  $a \to 0$. This can be checked in one-loop
perturbation theory and is a consequence of the mixing of the operator
$\bar h D_4 h$ with $\bar h  h$, as mentioned above. It has been argued
that it is possible to subtract the divergent term by computing the
coefficient of $\bar h h $  in perturbation theory \cite{lepage}.
Although it is true that with a hard (i.e. dimensional) ultraviolet
cut-off, such as the lattice spacing or the Pauli-Villars regulator,
the matrix elements of the bare operators have no renormalon
ambiguities, the subtraction of the power divergences using
perturbation theory reintroduces renormalons \cite{d2reno}. In other words, the
perturbation series for the power-divergent counterterms contain
renormalon ambiguities, which, as always, manifest themselves as
terms which are exponentially small in the coupling constant
eq. (\ref{eq:argu}). Thus the subtraction of power divergences has
to be performed non-perturbatively if the resulting matrix elements,
such as $\lb$, are to be unambiguous.

The matrix elements of the kinetic energy operator,
$\lambda_1$, also contain power divergent contributions. In this case,
the origin of the divergences is the mixing of  $\bar h \vec D^2 h$
with the operator $\bar h D_4 h$, with a  coefficient that diverges
linearly, and with the scalar density $\bar h  h$, with a quadratically
divergent  coefficient  \cite{mms}.  $\lambda_1$ determines the $1/m_Q$
corrections to
the heavy quark mass and hence enters many theoretical expressions of
weak decay factors. As in the case of $\labar$, the quadratic and
linear divergences of $\lambda_1$  must be subtracted
non-perturbatively.

The  numerical values  of $\labar$ and $\lambda_1$, presented in this
paper, have been obtained by using the non-perturbative method proposed
in ref. \cite{d2reno}. In that work, it has been shown that a
non-perturbative renormalisation prescription, which can be implemented
in lattice simulations, exists such as to avoid simultaneously both
power divergences and renormalon ambiguities,
in matrix elements and coefficient functions separately. In a theory
regulated by a dimensionful cut-off, it is
consistent not to perform the subtractions of the power divergent
terms at all, but to work with the bare operators and to compute the
coefficient functions (which will therefore contain powers of the cut-off)
in perturbation theory \cite{bigi}. In this case however, the matrix
elements in the effective theory are divergent in the ultra-violet
cut-off and depend on the regularization. Therefore
they cannot be interpreted as ``physical" quantities, in contrast to the
approach that we adopt here.

The linear divergence in $\labar$ is eliminated by a suitable
redefinition of the operator $\bar h D_4 h$. This definition corresponds
to the same normalization condition that one usually imposes in
perturbation theory. We require that the matrix element of  a
combination of  $\bar h D_4 h$ and $\bar h h$, $\bar h D^s_4 h=\bar h
D_4 h + \delta \overline{m} \bar h h$, is zero for  given  external
heavy quark states, in the Landau gauge: $\langle h( p_4=0)\vert  h
D^s_4 h  \vert h( p_4=0) \rangle = 0$ \footnote{ This requires
certain assumptions on the infrared behaviour of the heavy
quark propagator that will be discussed below, see also \cite{d2reno}.}.
Contrary to the perturbative
procedure, which reintroduces the renormalon ambiguities in the  matrix
elements of the subtracted operator, the non-perturbative
renormalization condition is unambigous, though prescription dependent,
and independent of the regularization procedure.
It is also quite natural in the sense that it allows a ``physical"
definition of $\labar$ that is finite and independent of the
ultraviolet cut-off. This procedure can be extended to the operators
appearing in higher orders of the $1/m_Q$ expansion.  Moreover it
allows the matching of the operators of the HQET to those in the full
theory (QCD) to be performed, via a combination of perturbative and
non-perturbative calculations, in such a way that the Wilson
coefficient functions are free of non-perturbative ambiguities at any
given order in the $1/m_Q$ expansion.

We show below that accurate results are obtained for the binding energy
to this order. Our best estimate is
\be \bar \Lambda= 190  \err{50}{30}\
  \hbox{MeV   }\, ,  \label{eq:vlb} \ee
where the error has been obtained by combining the statistical and
systematic errors, as will be discussed below. Our results show that,
as expected \cite{d2reno}, $\bar \Lambda$ is indeed independent of the
ultra-violet cut-off $a^{-1}$, within  reasonably small statistical and
systematic errors.

In order to remove the power divergences from the kinetic energy
operator, we have imposed on the relevant  operator a renormalisation
condition which corresponds to the ``physical"  requirement  $\langle
h(\vec p=0)\vert \bar h \vec D_s^2 h  \vert h(\vec p=0) \rangle = 0$,
where $\bar h \vec D_s^2 h$ is the subtracted kinetic energy operator
\cite{d2reno}.
This renormalisation condition, which will be explained in detail in
the next section, has been used to extract the values of the
mixing coefficients of the kinetic energy operator with the lower
dimensional ones
with a small statistical error. Unfortunately, after the  subtraction
of the power divergences, we were unable to obtain a precise value  for
$\lambda_1$, because of the large cancellations between the operator
matrix element and its counterterm. We  can only put a loose upper
bound
of $1$  GeV$^2$ on $\lambda_1$. Nevertheless, the results of this study
are so encouraging that  we are implementing this procedure on the
APE100 computer to perform a high statistics lattice calculation of
both $\labar$ and $\lambda_1$,  whose results will be published
elsewhere. We believe that the present results  demonstrate the
feasibility of the method proposed in ref. \cite{d2reno} to compute
quantities relevant in heavy flavour phenomenology. In this way, it
will still  be possible to use the HQET and the notion of a   ``pole"
mass,
now defined non-perturbatively,
which seemed to be ruined by the presence of the
renormalons.  Preliminary results of the present study can be found in
ref. \cite{mabie}.

The plan of the paper is the following. In sec. \ref{labardef} we
introduce the relevant formulae which define the non-perturbative
procedure for renormalising the operators $\bar h  D_4 h$ and $\bar h
\vec D^2 h$ \cite{d2reno}; in sec. \ref{numerical} we describe the
numerical calculation of $\labar$ and $\lambda_1$ and discuss the main
results of this study; in the conclusion we present the outlook for
future developments and applications of the method discussed in this
paper.

\section{Non-perturbative definition of $\labar$ and $\lambda_1$}
\label{labardef}

In this section we define the renormalisation prescription which we
will use to calculate ``physical" values of $\labar$ and $\lambda_1$.
The prescription involves imposing appropriate renormalisation
conditions on the quark matrix elements of the operators $\bar h D_4 h$
and $\bar h \vec D^2 h$, such that all their matrix elements are free
of power divergences \cite{d2reno}. Similar methods have been used for
light quark operators in refs. \cite{nperenor}--\cite{bernard}.

In numerical simulations, quark and gluon propagators can be computed
non-pertur\-batively by working in a fixed gauge, typically the Landau
gauge \cite{nperenor}--\cite{stella}.
The heavy quark propagator, at lowest order in $1/m_Q$, has the form
\be
 S(\vec x ,t)\, =\, \langle S(\vec x ,t\vert \vec 0,0) \rangle\, =\,
\delta(\vec x)\, \theta(t)\, \delta^{ij} \, A(t)\, \exp (-\lambda t)
\label{prop}
\ee
where $i,j$ are colour indices;
\beq S(\vec x ,t\vert \vec y,w)= \delta(\vec x -\vec y)
\, \theta(t-w)\,  \exp\Bigl(i \int_w^t A_0(t^\prime) dt^\prime \Bigr)
\label{cpline} \eeq
is the non-translationally invariant
propagator for a given gauge field configuration, computed in a given smooth
gauge, typically the Landau gauge, and
 $\langle \dots \rangle$ represents the average over the gauge
field configurations.
$A(t)$ is an unknown function of $t$, and we assume that
it decreases more slowly than an exponential at large times, specifically
we require that
\be\lim_{t \to \infty}
\frac{1}{a}  \ln \Bigl( \frac{A(t+a)}{A(t)} \Bigr)\sim
\lim_{t \to \infty}\frac{d}{dt}\ln{A(t)}  =0 .\label{eq:condition}\ee
Below we will show that, within the precision of our simulations, our
results for the heavy quark propagator are consistent with the
condition in eq. (\ref{eq:condition}). As will also be explained below,
the condition (\ref{eq:condition}) is not strictly required for the
definition and determination of $\lb$, since this can be done using
the values of the propagator at small times $t$. However in that case
it no longer has a direct interpretation as a binding energy, and our
preferred definition of $\lb$ does use the behaviour of the propagator
at large $t$.

The constant $\lambda$ in eq. (\ref{prop}) is linearly divergent in
$1/a$ and would correspond, in dimensional regularisation, to an
ultraviolet renormalon  in the effective heavy-quark propagator. Since
the linear divergence in $\lambda$ is due to the mixing of the operator
$\bar h D_4 h$ with the conserved scalar density operator $\bar h h$,
we can remove it by adding to the Lagrangian of the lattice HQET
\be
{\cal L}_{\rm eff} \,=\, \bar h(x)\, D_4\, h(x)
\label{eq:l0eff}
\ee
a counter-term of the form $\delta m\, \bar{h}(x)\, h(x)$. The HQET
Lagrangian then becomes
\be
{\cal L}^\prime_{\rm eff} \,=\, \frac{1}{1+ a \,\delta m}
\Bigl( \bar h(x)\, D_4\, h(x)\, +\,
\delta m\, \bar{h}(x)\, h(x)\Bigr),
\label{eq:l0effp}
\ee
where the factor $1/(1+a \, \delta m)$ has been introduced to ensure
the correct normalization of the heavy quark field $h$. With the action
${\cal L}^\prime_{\rm eff}$, the heavy-quark propagator is given by:
\be
 S^{ \prime}(\vec x ,t)\, =\,  \delta(\vec x)\, \theta(t)\, \delta^{ij} \,
A(t) \,  \exp{\left(\,- \left[\lambda + \frac{\ln (1+a \,\delta m )}{a}
\right]\, t\right)}\, .
\label{propp}  \ee
The mass counter-term  is defined by the  behaviour
of $S(\vec x ,t)$ at large values of the time
\beqn
 -\, \delta \overline{m}\, \equiv\, \frac{\ln(1+a \,\delta m )}{a}\, =\,
\lim_{t \to \infty}
  \frac{1}{a}\, \ln\left[\frac{{\rm Tr}\Bigl(S(\vec x ,t+a)\Bigr) }
{{\rm Tr}\Bigl(S(\vec x ,t)\Bigr)}
\right] \,= \nonumber  \eeqn \beqn \lim_{t \to \infty}
\left[  \frac{1}{a}\, \ln\left(\frac{A(t+a)}
{A(t)} \right) \,  -\lambda \right] \,= -\lambda \, ,
\label{ct}  \eeqn
where the traces are over the colour quantum numbers,
and we have
assumed the validity of the condition in eq. (\ref{eq:condition}).
Our numerical results, support the validity of this condition and
the use of eq. (\ref{ct}) is our preferred determination of
$\deltambar$.

We now define the renormalised binding energy by
\be
\labar\, \equiv\, {\cal E}\, -\, \delta \overline{m}\, ,
\label{lare}
\ee
which corresponds to the following relation between the meson and
the heavy quark mass
\be
M_{H}\, =\, m_{Q}\, +\, {\cal E}\, -\, \delta \overline{m}
\label{lare2}
\ee
$m_Q$ can be interpreted as a subtracted pole mass, and contains no
renormalon effects. A similar relation can be found in the case of a
heavy baryon.

The definition of $\delta \overline{ m}$ given in eq. (\ref{ct})
is not unique. A possible alternative definition would be, for example,
\beqn
-\, \delta \overline{m}(t^*)\, \equiv\,
  \frac{1}{a}\, \ln \left[\frac{{\rm Tr}\Bigl(S(\vec x ,t^*+a)\Bigr)}
 {{\rm Tr}\Bigl(S(\vec x
,t^*)\Bigr)}\right]  =  -\lambda\, +\, \frac{1}{a}\,
 \ln\left(\frac{A(t^*+a)} {A(t^*)} \right), \label{ct1}  \eeqn
where $t^*$ is a given time at which we perform the subtraction. The
corresponding definition of $\labar$, see eq. (\ref{lare}) above, will
clearly depend on the choice of $t^*$: $t^*$  parametrizes  the
renormalisation prescription dependence and can be considered as the
renormalisation point in coordinate space.
For physical matrix elements, the residual mass appears only through
the combination $m_{Q}\, -\, \delta\overline{ m}$, in such a way that
different choices of $\delta \overline{ m}$ are compensated by
different values of $m_{Q}$  \cite{fnl}.
The use of the propagator at small times, $t^*\Lambda_{QCD}\ll 1$, to
define $\delta \overline{ m}(t^*)$, and hence $\lb(t^*)$ does not
require any assumption about the behaviour of $A(t)$ at large times,
and in section \ref{labarlead} we present the results for $\lb(t^*)$
obtained in this way.

In addition to the non-perturbative contribution of $O(\Lambda_{QCD})$
to $\deltambar (t^*)$, there is a perturbative one proportional to
$1/t^*$ \cite{d2reno},
\be
-\deltambar_{{\rm pert}}(t^*)=-\frac{\alpha_sC_F}{4\pi}
\frac{\gamma_\psi}{t^*} + O(\alpha_s^2)
\label{eq:deltampert}\ee
where $\gamma_\psi$ is the one-loop contribution to the anomalous
dimension of the heavy quark field ($\gamma_\psi = -6$ in the Landau
gauge) and $C_F$ is the quadratic Casimir operator in the fundamental
representation ($C_F=4/3$).
Thus the definition of $\lb (t^*)$ defined,
at small times cannot readily be identified as a physical binding energy.
Nevertheless, computed values of $\lb (t^*)$ can be used to determine
standard short-distance heavy quark masses (such as the $\overline{{\rm
MS}}$ one) using perturbation theory (as will be explained in section
\ref{mbmsb} below). This, together with the fact that no assumption
about the infra-red behaviour of the heavy quark propagator is necessary,
is of fundamental importance.

\subsection{ Non-perturbative subtractions for $\lambda_1$}
\label{subl1}
The renormalised kinetic operator
$\bar h \vec D_S^2 h$,
free of power divergences, has the form
\begin{eqnarray}
\kkrinetic  &=& \kkinetic
- \frac{c_1}{a}\, \frac{1}{(1+a \, \delta m)}
\Bigl(\, \energy\, +\, \delta m\, \unity\, \Bigr)\,
\nonumber \\ &-&\, \frac{c_{2}}{a^{2}}\, \unity,
\label{eq:d2ren}
\end{eqnarray}
where the constants $c_{1}$ and $c_{2}$ are functions of the bare
lattice coupling constant $g_0(a)$. They have been computed in one loop
perturbation theory in ref.~\cite{mms}. Notice that we have preferred
to express $\bar h\vec D^2_S h $ in terms of the subtracted operator
which explicitly contains the residual mass $\delta m$. In this way we
can use the equations of motion of the Lagrangian ${\cal L}^\prime$
given in eq.~(\ref{eq:l0effp}).
This will prove useful below. The constant $c_{2}$ enters in the
renormalisation of the heavy quark mass. Therefore, it will contribute
to the relation between $M_H$, $m_Q$ and $\labar$ at order $1/m_Q$ (see
below). On the other hand, the constant $c_{1}$ contributes to the
renormalisation of the heavy quark wave-function and hence to the
renormalisation of all the operators containing a heavy quark field,
but not to the relation for the quark mass.

In order to eliminate the quadratic and linear  power divergences,
a possible  non-perturbative  renormalisation condition for
 $\bar h\vec D^2_S h $  is that its subtracted matrix element, computed
 for a quark at rest in the Landau gauge, vanishes
\be
\langle h(\vec p=0) \vert \bar h\vec D^2_S h  \vert h(\vec p=0) \rangle=0.
\ee
This is equivalent to defining the subtraction constants through
the relation (in the following we will work in lattice units, setting $a=1$)
\beq
\rho_{\vec{D}^{2}}(t)  = c_1\,+\,c_2\, t ,\label{eq:twenty}\eeq
where
\beqn \rho_{\vec{D}^{2}}(t) \equiv  \frac{
\sum_{\vec{x}}\, \langle \, S^{a\;\prime}(\vec x,t\vert\vec 0,0)\rangle}
{\sum_{\vec{x}}\,\langle S^{\prime}(\vec x,t\vert \vec 0,0)\rangle} =
\frac{\sum_{t^\prime=0}^{t}\,
\sum_{\vec{x},\vec{y}}\,   \langle \, S^{\prime}(\vec x,t\vert
\vec y,t^\prime)\,  \vec D^2_y(t^\prime)\, S^{\prime}(\vec y,t^\prime\vert
\vec 0,0)\, \rangle}
{\sum_{\vec{x}}\,\langle S^{\prime}(\vec x,t\vert \vec 0,0)\rangle}
\label{eq:c12}
\end{eqnarray}
By fitting the time dependence of $\rho_{\vec{D}^{2}}(t)$ to
eq.~(\ref{eq:twenty}), one obtains $c_1$ and $c_2$.

The heavy-quark propagator that enters in eq.  (\ref{eq:c12}) is the
subtracted  one, i.e. it is calculated with the action
(\ref{eq:l0effp}) instead of  (\ref{eq:l0eff}). We now  demonstrate
that  $\rho_{\vec{D}^{2}}(t)$  can be expressed  in terms of
unsubtracted propagators only:
\[  \rho_{\vec{D}^{2}}(t) =
\frac{\sum_{t^\prime=0}^{t}\,\sum_{\vec{x},\vec{y}}\,
 \langle(S^{ \prime}(\vec x,t\vert \vec y,t^\prime)\,
\vec D^2_y(t^\prime)\, S^{ \prime}(\vec y,t^\prime\vert \vec 0,0)\rangle)}{
\sum_{\vec{x}}\,\langle S^{ \prime}(\vec x,t\vert \vec 0,0)\rangle}\, =
\]
\[
\frac{\sum_{t^\prime=0}^{t}\,\sum_{\vec{x},\vec{y}}\,
 \langle(S(\vec x,t\vert \vec y,t^\prime)
e^{\delta \overline{m}\,(t-t^\prime)}\,
\vec D^2_y(t^\prime)\, S(\vec y,t^\prime\vert \vec 0,0)\,
e^{\delta \overline{m}\, t^\prime}\rangle)}{
\sum_{\vec{x}}
\,\langle S(\vec x,t\vert \vec 0,0)\, e^{\delta \overline{m}\, t}
\rangle}\, = \]
\be
\frac{\sum_{t^\prime=0}^{t}\sum_{\vec{x},\vec{y}}\,
 \langle(S(\vec x,t\vert \vec y,t^\prime)
\, \vec D^2_y(t^\prime)\, S(\vec y,t^\prime\vert \vec 0,0)\rangle)}{
\sum_{\vec{x}}\,\langle S(\vec x,t\vert \vec 0,0)\rangle}
\ee
Notice that this argument holds for any operator which does not contain
a time derivative.

For some important applications it is only the constant $c_2$ which is
required. $c_2$ can also be determined  directly by eliminating the sum over
$t^\prime$ in eq. (\ref{eq:c12}):
\be
c_2 =\rho_{\vec{D}^{2}}(t^\prime,t) =\frac{\sum_{\vec x,\vec y}\,
 \langle(S^{ \prime}(\vec x,t\vert \vec y,t^\prime)
\, \vec D^2_y(t^\prime)\, S^{ \prime}(\vec y,t^\prime\vert \vec 0,0)\rangle)}{
\sum_{\vec{x}}\,\langle S^{ \prime}(\vec x,t\vert \vec 0,0)\rangle}
\label{eq:c2tp}\ee
for $t^\prime\neq 0, t$.

The relation between the mass of the meson and the mass  of the quark
to order $1/m_{Q}$ is then given by
\be
M_{H}\, =\, m_{Q}\, +\, {\cal E}\, -\, \delta \overline{m}\, - \,
\left( 1- \frac{\alpha_s}{4\pi}X_{\vec D^2_S}\right)
\left(\frac{\lambda_1^{{\rm bare}}
- c_{2}}{2\, m_{Q}}\right) +O(\frac{1}{m_Q^2}),
\label{eq:nextor}
\ee
where $\lambda_1^{{\rm bare}}=\langle B
\vert \bar h \vec{D}^{2} h \vert B \rangle/(2 M_B)$.
$\lambda_1^{{\rm bare}}$ can
be determined from a computation of two- and three-point correlation
functions in the standard way. Consider the meson three-point
correlation function (the extension of this discussion to baryons
is entirely straightforward)
\beq
C_{\vec D^2}(t^\prime, t) = \sum_{\vec x, \vec y}
\langle 0|J_\Gamma(\vec x, t)\ \bar h(\vec y,t^\prime)\vec D_y^2 h(\vec
y,t^\prime)
J^\dagger_\Gamma(\vec 0,0)|0\rangle
\label{eq:cd2s}\eeq
For sufficiently large values of $t^\prime$ and $t-t^\prime$
\beq
C_{\vec D^2}(t^\prime, t)\ \rightarrow Z^2 \, \lambda_1^{{\rm bare}}\,\exp
\left(-({\cal E}-\delta \overline{m})t\right)\, .
\label{eq:cd2sasymp}\eeq
A convenient way to extract $\lambda_1^{{\rm bare}}$ is to
consider the ratio
\beq
R(t^\prime, t)=\frac{C_{\vec D^2}(t^\prime, t)}{C(t)}\rightarrow
\lambda_1^{{\rm bare}}
\label{eq:deltaasymp}\eeq
As usual $\lambda_1^{{\rm bare}}$ must be evaluated in an interval in which
$R(t^\prime, t)$ is independent of the times $t^\prime$ and $t$, so that the
contribution from  excited states and contact terms can be neglected.

The term proportional to $X_{\vec D^2_S}$ in eq. (\ref{eq:nextor}) is
absent in continuum formulations of the HQET, and is a manifestation
of the lack of reparametrisation invariance in the lattice version. It
has been calculated in ref. \cite{mms}.
Notice that only the constant $c_{2}$ enters the eq. (\ref{eq:nextor}) because
$c_{1}$ is eliminated by using the equations
of motions. $c_{1}$  only modifies the wave function renormalisation of
the heavy quark, thus  contributing to  the $O(1/m_Q)$ corrections
of the hadronic matrix elements.

\section{Numerical implementation of the renormalisation procedure}
\label{numerical}

As explained in the previous section, the determination of $\lb$ and
$\lambda_1$ requires the computation of the quark propagator and
matrix elements between quark states in a fixed gauge (in order to
obtain the subtracted operators), as well as the evaluation of matrix
elements between hadronic states. We have obtained our results using
five independent numerical simulations, whose main parameters
are given in table \ref{tab:sets}.
\begin{table} \centering
\begin{tabular}{||c|c|c|c||}
\hline
\hline
simulation & volume & $\beta$& Number of configurations
\\ \hline  \hline
 \underline{set A}& $16^{3}\times 32$ &$6.0$&
$36$ \\ \hline
  \underline{set B}& $16^{3}\times 32$& $6.0$&
$300$  \\ \hline
 \underline{set C} &$20^{3}\times 32$ &$6.2$&
$50$ \\ \hline
  \underline{set D} & $18^{3}\times 64$& $6.0$&
$210$ \\ \hline
  \underline{set E}& $18^{3}\times 64$& $6.2$&
$420$ \\ \hline \hline
\end{tabular}
\caption{\it{Parameters of the numerical simulations, the
 results of which have been used for the present study.}}
\label{tab:sets}
\end{table}

Our best value of the subtracted binding energy,
$\bar \Lambda= {\cal E} - \delta \overline{m}$, has been determined by
combining the values of $\delta \overline{m}$ obtained using
\underline{set B} and  \underline{set C} with the calculation of ${\cal
E}$ performed by the APE collaboration at $\beta=6.0$, \underline{set
D}, and $6.2$, \underline{set E} \cite{alltonw,allton}. $\cal E$ had
been determined using the SW-Clover fermion action for the light
quarks. The calculations were performed at several masses of the light
quark, so that extrapolations to the chiral limit are possible. We also
present the results for the subtraction constants $c_1$ and  $c_2$
obtained with \underline{set B} and \underline{set C}.

So far we have only computed $\lambda_1^{{\rm bare}}$ using
\underline{set A}. Again, for the light quarks the improved SW-Clover
action \cite{clover} was used in the quenched approximation. These
exploratory calculations were performed at one value of the mass of the
light quark, $\kappa = 0.1425$, for which the  mass of the
corresponding ``pion" is about 900\,MeV. The details of the simulation
can be found in refs. \cite{nperenor,msv}. Preliminary results for both
$\lb$ and $\lambda_1$ for mesons, evaluated using this dataset have
been presented in  ref.~\cite{mabie}.

All the errors have been computed  with the jacknife method by
decimating  one configuration at a time (\underline{set A}  and
\underline{set C}) or five configurations at a time  (\underline{set
B}). The error on ${\cal E}$ was computed with the jacknife method also
and we refer the reader to refs. \cite{alltonw,allton} for details.

\subsection{Determination of the residual mass $\delta \overline{m}$}
\label{mbmsb}
A possible lattice expression for the forward heavy-quark propagator,
to leading order in the heavy quark mass, is given by
\be
S(x \vert 0)\, =\,\delta(\vec{x})\, \theta(x^4)\,
{\cal P}_{\vec x}(x^{4}\, \vert \, 0 ) \label{ls0}\ee
where ${\cal P}_{\vec x}\left(x^{4}\, \vert\, y^{4}\right)$ is  the
lattice path ordered exponential from $(\vec{x}, y^{4})$ to $(\vec{x},
x^{4})$, cf. eq.~(\ref{cpline}), usually called ``P-line",
\begin{eqnarray}
{\cal P}_{x}(x^{4}\, \vert \,y^{4}) &=&
\prod_{n=1}^{\left[\frac{x^{4}-y^{4}}{a}\right]} \,
U^{\dag}(\vec{x},x^{4}\, -\, n\, a) , \,\,\,\,\,\, x^4 > y^4 \nonumber
\\ {\cal P}_{x}(x^{4}\, \vert \,y^{4}) &=&
1\,\,\,\,\,\, x^4 = y^4
\label{pline}
\end{eqnarray}
This propagator corresponds to the following choice for the covariant
time derivative, $D_4\,f(t)=1/a\,(f(t)-U^{\dagger}_4(t-a)f(t-a)\,)$.

In order to reduce the statistical noise,  we have computed, in the
lattice Landau gauge, the quantity
\be
S_{H}(t)\, =\, \frac{1}{3 V}\, \sum_{\vec{x}}\,
\langle {\rm Tr}\left[\, {\cal P}_{\vec{x}}(x^4=t\, \vert\, 0 )\, \right]
\rangle ,\ee
where the trace is over the colour indices and $V$ denotes the spatial
volume of our lattice.
It is this averaged propagator $S_{H}(t)$, which has been used in the
computations below.

There is a subtle
point that we would like to discuss briefly.  It can be demonstrated
that $O(a)$ effects in heavy-light operator matrix elements between
physical states are cancelled by improving  the light quark propagators
only \cite{clover2}. On the other hand, in order to improve off-shell
matrix elements, which is the case when renormalising the operators
between quark states, it is necessary to use an improved version of the
heavy quark propagator in the effective theory, for example
\be
{\cal P}^{I}_{x}(x^{4} \, \vert \,y^{4})\, =\, \left[\, 1 \, -\,
\left(\frac{1}{3}\right)^{y^{4}-x^{4}+1}\, \right]\,
{\cal P}_{x}(x^{4}\, \vert \,y^{4}) \, . \label{ipl}
\ee
Notice that the improved P-line tends very rapidly to  the unimproved
one as $x^{4}-y^{4}$ increases.
The propagator in eq.(\ref{ipl}) corresponds to the following time
derivative $D_4f(t) =1/a\,(3/2f(t) -2 U^{\dagger}(t-a)f(t-a) +
1/2U^{\dagger}(t-a) U^{\dagger}(t-2a)f(t-2a)\,)$. It is also possible to
add a residual mass term to the heavy quark action in such a way that
it modifies the propagator (\ref{ipl}) by an exponetial in time (up to
an overall normalisation factor). Such as mass term takes the form
$3/2\,(1/\lambda-1)\bar h(t)h(t) + 1/2\, (\lambda-1)\bar h(t)h(t-2a)$.
In the following, when discussing the improved heavy quark propagator,
we will implicitly assume that the mass term is of this form.

To determine the residual mass, we have to compute the effective mass of
the propagator $S_{H}(t)$, defined by
\be
a \,  \delta \overline{m}(t)\,
=\, -\, \ln\left(\frac{S_{H}(t+a)} {S_{H}(t)}\right)
\label{effmas}
\ee
\begin{figure}
\vspace{9pt}
\begin{center}\setlength{\unitlength}{1mm}
\begin{picture}(160,100)
\put(0,-40){\special{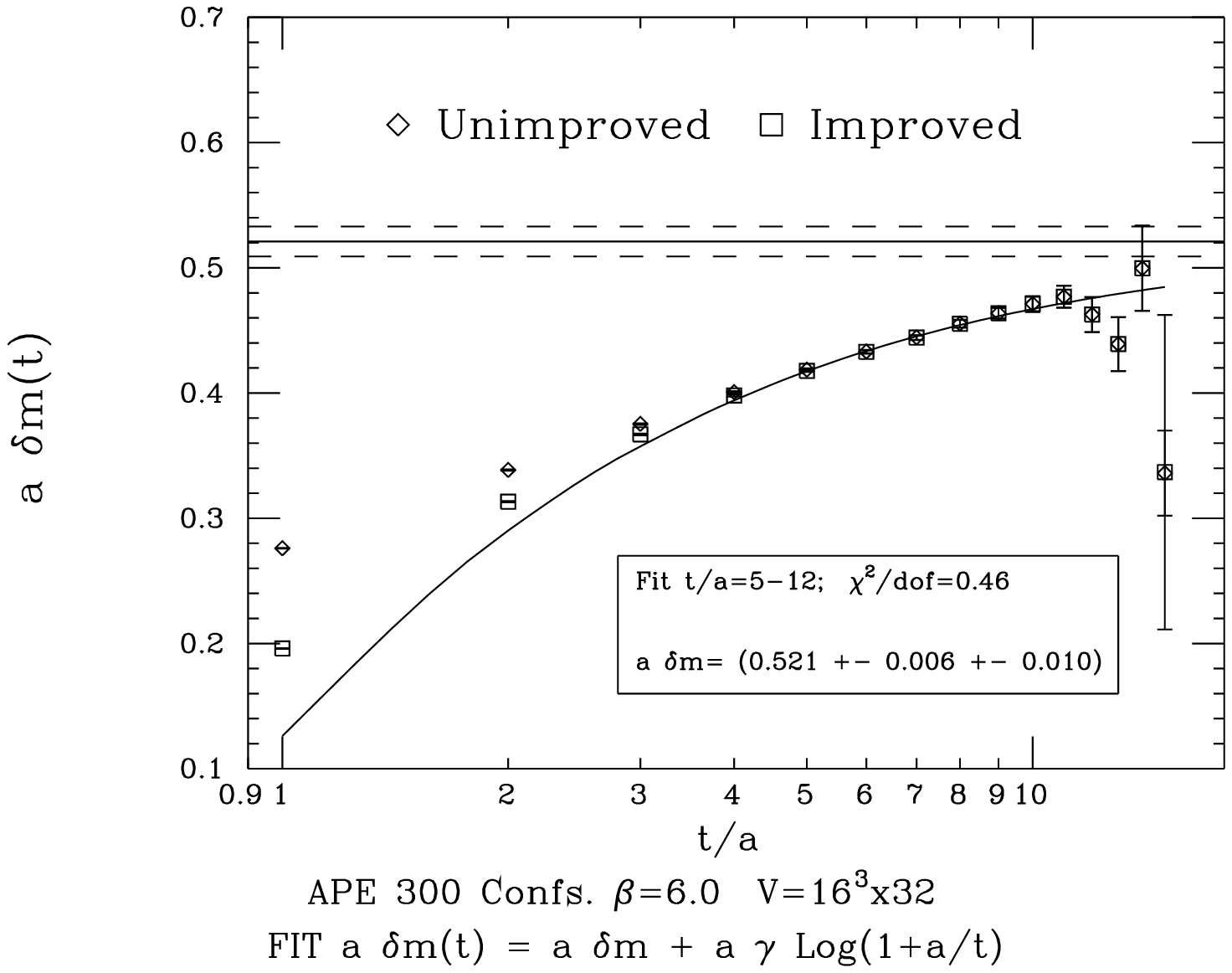}}
\end{picture}
\end{center}
\caption{\it{Effective mass of the heavy-quark propagator $S_{H}(t)$, at
$\beta=6.0$, as a
function of the time. The curve represents a  fit of the numerical results
(in the improved case) to the expression given in eq. (34).}}
\label{fig:fit60}
\end{figure}
\begin{figure}
\vspace{9pt}
\begin{center}\setlength{\unitlength}{1mm}
\begin{picture}(160,100)
\put(0,-30){\special{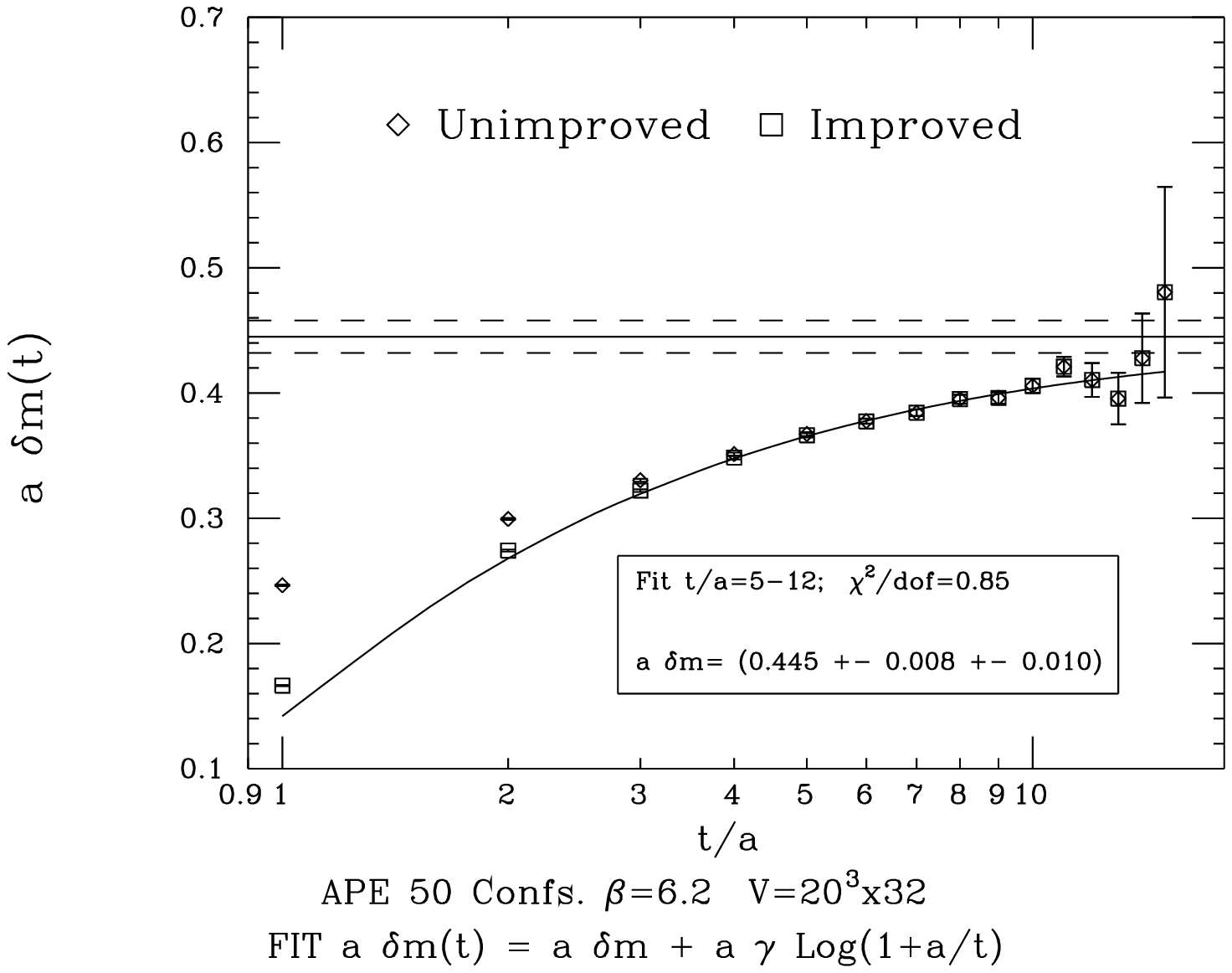}}
\end{picture}
\end{center}
\caption{\it{Effective mass of the heavy-quark propagator $S_{H}(t)$, at
$\beta=6.2$, as a
function of the time. The curve represents a  fit of the numerical results
(in the improved case) to the expression given in eq.
(\protect\ref{eq:fit2}).}}
\label{fig:fit62}
\end{figure}
In figs. \ref{fig:fit60} (from \underline{set B}) and \ref{fig:fit62}
(from \underline{set C}),  we present the values of
$\delta \overline{m}(t)$ for the
improved and unimproved propagators as a function of $t/a$.
The effective mass is indistinguishable in the two (improved and unimproved)
cases,
for $t/a > 4$--$5$. Thus, in order to minimize lattice artefacts,
 we have only used  the results obtained for $t/a \ge 5$.
 Inspired
by  the results of one-loop perturbation theory \cite{d2reno},
 we made a fit to $\delta \overline{m}(t)$ using the expression
\be a \, \delta \overline{m}(t)  = a \, \delta \overline{m} + \gamma\frac{a}{t}
\label{eq:fit}
\ee
In order to mimic higher order effects, we have also used different
expressions to fit $\delta \overline{m}(t)$, e.g.
\be a \, \delta \overline{m}(t)  =a \,  \delta \overline{m} + \gamma^\prime
\ln \Bigl( \frac{t+a}{t}\Bigr)  \label{eq:fit1}
\ee
or
\be a \,  \delta \overline{m}(t)  = a\, \delta \overline{m} -
{\gamma^{\prime\prime}}\ln\Bigl(\frac{\alpha_s[{\cal K}/(t+a)]}
{\alpha_s[ {\cal K}/t]}
\Bigl) \to
a \, \delta \overline{m} +
\gamma^{\prime\prime}\ln\Bigl(\frac{\ln[(t+a)]+{\cal C}}{\ln[t]+ {\cal C}}
\Bigl),  \label{eq:fit2}
\ee
and changed the interval of the fits in order to check the stability
of the determination of $ \delta \overline{m}$.
 In eqs. (\ref{eq:fit})--(\ref{eq:fit2}),
$\delta \overline{m}$, $\gamma, \dots, \gamma^{\prime\prime}$ and ${\cal C}$
are free parameters of the fit.
The curves shown in fig. \ref{fig:fit60} and \ref{fig:fit62}
 correspond to  fits of the improved
heavy quark propagator to eq. (\ref{eq:fit1}),
 in the interval  $5 \le t/a \le 12$.
\begin{table} \centering
\begin{tabular}{||c|r|r|r|r|r|r||}
\hline
\hline
\multicolumn{7}{||c||}{Subtraction constant $a \delta \overline{m}$
  at $\beta=6.0$}\\
\hline
\hline
Fit&\multicolumn{1}{c|}{$t=4-12$}
&\multicolumn{1}{c|}{$t=5-12$}&\multicolumn{1}{c|}{$t=5-14$}&
\multicolumn{1}{c|}{$t=6-14$}&
\multicolumn{1}{c||}{$t=7-14$}&\multicolumn{1}{c||}{$t=8-14$}\\
\hline
\hline
eq. (33)&$0.507(4)$&$0.515(6)$&$0.515(6)$&
$0.52(1)$&$0.52(2)$&$0.51(3)$\\
$\chi^2$/dof&$1.50$&$0.56$&$0.86$&
$0.84$&$0.96$&$1.05$\\
\hline
eq. (34)&$0.515(4)$&$0.521(6)$&$0.521(6)$&
$0.53(1)$&$0.53(2)$&$0.51(3)$\\
$\chi^2$/dof&$0.97$&$0.46$&$0.81$&
$0.85$&$0.98$&$1.05$\\
\hline
eq. (35)&$0.513(4)$&$0.521(6)$&$0.520(6)$&
$0.51(3)$&$0.51(1)$&$0.50(2)$\\
$\chi^2$/dof&$1.20$&$0.56$&$0.93$&
$0.97$&$1.10$&$1.28$\\
\hline
\hline
\end{tabular}
\caption{\it{Numerical values of the constant $a \delta \overline{m}$ found by
using
the results of \underline{set B}, at $\beta=6.0$. The results are  from
 several fits in different time intervals. We also give the uncorrelated
$\chi^2$/dof in the different cases. The numbers given in this table refer
to the improved heavy quark propagator only.}}
\label{tab:fit60}
\end{table}

%%%%%%%%%%%%%%%%%%%%%%%%%%%%%%%%
\begin{table} \centering
\begin{tabular}{||c|r|r|r|r|r|r||}
\hline
\hline
\multicolumn{7}{||c||}{Subtraction constant $a \delta \overline{m}$
  at $\beta=6.2$}\\
\hline
\hline
Fit&\multicolumn{1}{c|}{$t=4-12$}
&\multicolumn{1}{c|}{$t=5-12$}&\multicolumn{1}{c|}{$t=5-14$}&
\multicolumn{1}{c|}{$t=6-14$}&
\multicolumn{1}{c||}{$t=7-14$}&\multicolumn{1}{c||}{$t=8-14$}\\
\hline
\hline
eq. (33)&$0.437(5)$&$0.441(7)$&$0.440(8)$&
$0.45(1)$&$0.46(2)$&$0.45(2)$\\
$\chi^2$/dof&$0.50$&$0.93$&$0.80$&
$0.75$&$0.70$&$0.83$\\
\hline
eq. (34)&$0.442(5)$&$0.445(8)$&$0.445(8)$&
$0.45(1)$&$0.46(2)$&$0.46(3)$\\
$\chi^2$/dof&$0.40$&$0.85$&$0.74$&
$0.73$&$0.70$&$0.83$\\
\hline
eq. (35)&$0.443(6)$&$0.445(8)$&$0.445(8)$&
$0.45(1)$&$0.45(5)$&$0.46(3)$\\
$\chi^2$/dof&$0.50$&$1.02$&$0.85$&
$0.85$&$0.84$&$1.05$\\
\hline
\hline
\end{tabular}
\caption{\it{Numerical values of the constant $a
 \delta \overline{m}$ found by using
the results of \underline{set C}, at $\beta=6.2$. The results are  from
 several fits in different time intervals. We also give the uncorrelated
$\chi^2$/dof in the different cases. The numbers given in this table refer
to the improved heavy quark propagator only.}}
\label{tab:fit62}
\end{table}
%%%%%%%%%%%%%%%%%%%%%%%%%%%%%%%%%%%%%%%%%%%%
{}From the different results obtained by varying the fitting functions and the
time intervals, see tables \ref{tab:fit60}
and \ref{tab:fit62},  we quote
\be a \,\delta \overline{m} = 0.521 \pm 0.006 \pm 0.010 \hbox{ at }
 \beta=6.0 \label{dbm60} \ee
\be a \,\delta \overline{m} = 0.445 \pm 0.008 \pm 0.010 \hbox{ at }
 \beta=6.2 \label{dbm62} \ee
where in both cases the first error is statistical, and the second is
an estimate of the systematic uncertainty, based on the spread of results
obtained using different time intervals and fitting functions.

The determination of the mass counter-term at fixed $t=t^*$, requires
no fitting, and the results obtained using \underline{set B} and
\underline{set C} are presented in table \ref{tab:lt} below.

\subsection{Determination of $\labar$}
\label{labarlead}

We are now ready to present our prediction for $\labar$. In order to
evaluate the subtracted  $\labar$, we have used the results of the high
statistics calculations of ${\cal E}$ given in
refs.~\cite{alltonw,allton} (\underline{set D} and \underline{set E}).
We will also make use of the results obtained by using the standard
Wilson action, on a $18^3 \times 64$ lattice, at $\beta=6.0$, with a
statistical sample of $200$ configurations~\cite{alltonw}.

In order to obtain $\bar \Lambda$ we have used:
\begin{itemize}
\item $\delta \overline{m}$ from eqs. (\ref{dbm60}) and (\ref{dbm62});
\item  the SW-Clover determination of ${\cal E}$ of the APE collaboration,
$a {\cal E}=0.61 \pm 0.01$ at $\beta=6.0$
and $a {\cal E}=0.52 \pm 0.01$ at $\beta=6.2$~\cite{alltonw,allton};
\item $a^{-1}(\beta=6.0)=2.0 \pm 0.2$ GeV and
$a^{-1}(\beta=6.2)=2.9 \pm 0.3$ GeV.
The calibration of the lattice spacing in quenched simulations
typically  has an uncertainty of $O(10\%)$, depending on the physical
quantity which is used to set the scale. We take these results as a fair
representation of the spread of possible values.
\end{itemize}
We then find
\beqn
\lb & = &{\cal E}-\delta \overline{m} \,=\,
180 \pm 35\ {\rm MeV}\ {\rm at}\ \beta =6.0
\label{eq:lb6.0}\\
\lb & = & {\cal E}-\delta \overline{m} \,=\,
220 \pm 55\ {\rm MeV}\ {\rm at}\ \beta =6.2
\label{eq:lb6.2}\eeqn
where the statistical errors have been combined in quadrature with
those due to the uncertainty in the lattice spacing.

Within the uncertainties, the results in eqs.~(\ref{eq:lb6.0}) and
(\ref{eq:lb6.2}) are compatible with the expected independence of
$\bar\Lambda$ of the lattice spacing. Given the intrinsic uncertainty
in the value of the lattice spacing in quenched simulations it is
difficult however, to check this more precisely.
Indeed, using different physical quantities to set the scale can
increase or decrease the difference in the central values of $\lb$ at
$\beta=6.0$ and 6.2. For example, using the string tension to set the
scale one finds $a^{-1}(\beta=6.0)= 1.88$ GeV and
$a^{-1}(\beta=6.2)=2.55$ GeV,
giving $\lb = 170\pm 30$~MeV at $\beta=6.0$ and  $\lb =190\pm
40$ MeV at $\beta$=6.2, whereas using the mass of the  $\rho$-meson to
set the scale the APE collaboration finds $a^{-1}(\beta=6.0)=1.95 \pm
0.07$ GeV and   $a^{-1}(\beta=6.2)=3.05 \pm 0.20$ GeV
\cite{alltonw,allton},
 which corresponds to $\lb = 176\pm 30$ MeV at $\beta=6.0$ and $\lb
=228\pm 50$ MeV at $\beta=6.2$~\footnote{Notice that, using the mass of
the  $\rho$-meson,  the UKQCD collaboration found $a^{-1}(\beta=6.2)
\sim 2.7(1)$~GeV~\cite{ukqcdmrho}, corresponding to $\bar
\Lambda=203\pm 45$\,MeV.}. Nevertheless in both cases the results are
compatible at the two values of $\beta$.

Assuming that $\bar \Lambda$ is indeed constant in $a$, we combine the
results in eqs. (\ref{eq:lb6.0}) and (\ref{eq:lb6.2}) to obtain
\be \bar \Lambda= (190 \pm 30 ) \hbox{ MeV }\, .\label{lbres1} \ee
Before quoting our final result, we need to estimate the discretisation
error.

{}From a comparison of the values of $\delta \overline{m}$ obtained with
the improved and unimproved heavy quark propagators, we believe that
discretisation effects are negligible for this quantity. Indeed
discretisation errors in quantities which only depend on the gauge
fields are of $O(a^2 \,\Lambda_{{\rm QCD}}^2)$, when evaluated using
the Wilson gauge action. However, in the computation of $\lb$ (and
$\lambda_1$), correlation functions which contain the light quark
propagator are evaluated, and with the SW-Clover and Wilson fermion
actions this introduces errors of $O(\alpha_s a \, \Lambda_{{\rm
QCD}})$ and $O(a \, \Lambda_{{\rm QCD}})$ respectively.
Notice that these effects are  formally larger than the  higher order
$1/m_Q$ corrections to $\bar \Lambda$ (we work in the approximation
$a^{-1} \ll m_Q$). To obtain an estimate of the size of the
discretisation errors, we compare ${\cal E}$ obtained with the standard
Wilson action and the SW-Clover action at the same value of $\beta$,
$\beta=6.0$.
In the Wilson case, by working  at four different values of the light
quark mass, the bare binding energy, extrapolated to  the chiral limit
in the light quark mass, was found   to be $a \, {\cal E}_W=0.608(8)$.
In the SW-Clover case, by working  at three different values of the
light quark mass,  the result for the bare binding energy, extrapolated
to the chiral limit in the light quark mass, was found to be $a \,
{\cal E}_{SW}=0.616(4)$  \footnote{ As a check of our calculations with
\underline{set A}, we have verified that on these configurations, at
the value of the mass where we have computed the light quark propagator
($K=0.1425$), our results for ${\cal E}$ agree with those of refs.
\cite{alltonw,allton}.}. The difference between the central values obtained
with the two actions ${\cal E}_{SW}-{\cal E}_W=(0.616-0.608)\, a^{-1}
\sim 16$~MeV. We deduce that +20\,MeV is a reasonable estimate of the
discretisation error in the determination  of $\lb$. We therefore quote
as our final result for $\lb$
\be \bar \Lambda= 190 \err{50}{30} \hbox{ MeV }\, .\label{lbres} \ee

The prediction given in eq. (\ref{lbres}) can be compared with other
results that have been presented in the literature. In perturbation
theory one finds
\be a \,\delta \overline{m}_{{\rm pert}}= \frac{\alpha_s}{3} \int\frac{d^3 q}
{(2 \pi)^2} \,
\Bigl( \frac{1}{\sum_{i=1}^3 \sin^2(q_i/2)} \Bigr)
=2.12 \times \alpha_s \label{eq:pert1} \ee
By using values for $\alpha_s$  which are commonly proposed in the
literature for the ``boosted" coupling \cite{lm,fnal},
$\alpha_s=0.13$--$0.18$  at $\beta=6.0$, eq. (\ref{eq:pert1}) would
give $ a \,\delta \overline{m}_{{\rm pert}}=0.28-0.38$. Thus, even in
the most favourable case, $\delta \overline{m}_{{\rm pert}}$ is about
$280$ MeV smaller (i.e. $\bar \Lambda_{{\rm pert}}$ is  about $280$ MeV
larger) than our non-perturbative determination.

In ref. \cite{fnal}, the bare binding energy ${\cal E}$ has been
determined, using the Wilson action for light quarks,
on a variety of lattice volumes and at several values
of $\beta$, $\beta=5.7,5.9,6.1$ and $6.3$. The results are consistent
with a linear dependence
\be a\, {\cal E}(a) = {\cal E}_0 + a \, \labar_{{\rm FNAL}}\ee
where ${\cal E}_0$ and $\labar_{{\rm FNAL}}$ are parameters of the fit,
${\cal E}_0=0.351(14)$ and $\labar_{{\rm FNAL}}=0.481(25)$~GeV. The
value of ${\cal E}_0$ is consistent with  $a \,\delta \overline{
m}_{{\rm pert}}$ computed using an ``effective" $\alpha_s=0.166$ (this
value may be considered as an average of the values of the strong
coupling constant on the points in $\beta$ where ${\cal E}$ has been
computed).  On the other hand,  the value of the ``finite" binding
energy $\labar_{{\rm FNAL}}$ is about $300$ MeV larger than
ours \footnote{Given the presence of terms of $O(a)$ the stability of
the results with respect to a quadratic fit
of the form $a\, {\cal E}(a) = {\cal E}_0 +  \labar_{{\rm FNAL}}a\,+
{\cal E}_2 a^2$, where ${\cal E}_2$ is a constant, remains to be
checked.}.
Our interpretation is that, up to possible $O(a)$ effects, the two
determinations differ because of the finite non-perturbative
contribution of $O(\Lambda_{{\rm QCD}})$ that has been subtracted only
in our case. Using the definition of ref. \cite{fnal} however, it is
not clear how to match the full and the effective theories, since their
definition includes non-perturbative, uncalculable effects.

A further demonstration of the existence of the non-perturbative
effects is provided by the comparison of $a\delta  \overline{ m}$ and
$a^{-1}$ at  $\beta=6.0$ and $6.2$. In the  absence of non-perturbative
terms of  $O(\Lambda_{{\rm QCD}})$, i.e. if $\delta \overline{ m}$ is
given only by the linearly divergent contribution,  we should find
$R_m\equiv a\delta  \overline{ m}(\beta=6.0)/a\delta  \overline{
m}(\beta=6.2) \simeq R_{\alpha_s}=
\alpha_s(\beta=6.0)/\alpha_s(\beta=6.2)$. Numerically we find
$R_m=1.16(4)$ to be compared with $R_{\alpha_s}=1.03-1.06$: $1.03$ is
simply $6.2/6.0$;  $1.06$ has been estimated  from $R_{\alpha_s^{{\rm
eff}}}$, where $\alpha_s^{{\rm eff}}= \alpha_s^{{\rm latt}}/\langle
{\LARGE \Box} \rangle$ with $\alpha_s^{{\rm latt}}=(6/\beta)/(4\pi)$
and $\langle{\LARGE \Box} \rangle$ is the expectation value of the
plaquette.

We now present the results for $\lb$ defined at a fixed value of $t^*$
($\lb(t^*)$). In order to be able to use perturbation theory to
determine values corresponding to standard short distance definitions
of the heavy quark mass, $t^*$ must be chosen to be sufficiently small.
In table $\ref{tab:lt}$ we present the results for the mass counterterm
$\delta \overline{m}(t^*)$ at both $\beta$ = 6.0 and 6.2, obtained using
the configurations of \underline{set B} and \underline{set C}
respectively, for small values of $t^*$. We then combine the results
for  $\delta \overline{m}(t^*)$ with those for ${\cal E}$ obtained by
the  APE collaboration (\underline{set D} and \underline{set E}) to
obtain  $\lb(t^*)$. These results for $\lb(t^*)$ will be used in
section
\ref{subsec:mbar} to determine the $\overline{MS}$ mass.

\begin{table} \centering
\begin{tabular}{||c|c|c|c|c||}
\hline
\hline
{$\beta$} &
{$t^*/a$}&$a\delta\overline{m}(t^*)$&{$\pi/t^*$ (GeV)}
&{$\bar \Lambda(t^*)$ (MeV)}\\
\hline
\hline
$6.0$ &$3$& $0.3670(6)$ &$2.1 \pm 0.2$ &$ 490 \pm 20 \pm 50$ \\
$$ &$4$   & $0.3980(8)$&$1.6 \pm 0.2$ &$420 \pm 20 \pm 40$ \\
$$ &$5$   & $0.4177(9)$&$1.3 \pm 0.1$ &$390 \pm 20 \pm 40$ \\
$$ &$6$   & $0.4328(13)$&$1.0 \pm 0.1$ &$350 \pm 20 \pm 40$ \\
\hline
% $6.2$&$3$ &$3.0 \pm 0.3$ &$ 570 \pm  30 \pm 60$ \\
$6.2$&$4$ &$0.3484(13)$ &$2.3 \pm 0.2$ &$ 500 \pm 30\pm 50$ \\
$$&$5$ & $0.3663(16)$ &$1.8 \pm 0.2$ &$ 450 \pm 30 \pm 50$ \\
$$&$6$ & $0.3773(20)$  &$1.5 \pm 0.2$ &$ 410 \pm 30 \pm 40$ \\
$$&$7$ & $0.3842(27)$  &$1.3 \pm 0.1$ &$ 390 \pm 30 \pm 40$ \\
\hline
\hline
\end{tabular}
\caption{\it{Results for $\bar \Lambda(t^*)={\cal E}-\delta
\overline{m}(t^*)$ for different normalization times $t^*$,
using the results from \underline{set B}--\underline{set E}.
The first error on $\bar \Lambda(t^*)$ is obtained by combining
 the errors on ${\cal E}$ and $\delta
\overline{m}(t^*)$
in quadrature; the second error (and the error on $\pi/t^*$)
comes from the calibration of the lattice spacing.}}
\label{tab:lt}
\end{table}

We end this subsection with an obvious but important remark. $\bar
\Lambda$ can be defined in many different ways, which correspond to
different renormalization prescriptions for the renormalized $\bar h
D_4 h$ operators. The presentation of results or bounds for
$\bar\Lambda$ must therefore be accompanied by the definition of the
prescription to which they correspond.
\subsection{The $\overline{MS}$ mass of the $b$-quark}
\label{subsec:mbar}
We now give the relevant formulae necessary to match the subtracted
mass of the quark $m_Q^S$ to the running mass $\overline{m}_Q$,
computed in the $\overline{MS}$ scheme at the scale
$\mu=\overline {m}_Q$. We introduce
the following quantities
\beq m_Q^S(t^*)= M_H-\bar \Lambda(t^*)=M_H- {\cal E}+
\delta \overline{m}(t^*)\, , \eeq
\beq \overline{C}_m(t^*)=1 -\frac{4\alpha_s(
\overline {m}_Q)}{3 \pi} - \frac{1}{\overline{m}_Q}
\left( \delta \overline{m}(t^*)
-\alpha_s(a) \frac{X}{a} \right) \, .\eeq
It is straightforward  to show, that, at order $\alpha_s$, the relation
between $\overline{m}_Q$ and $m_Q^S(t^*)$ is given by
\beq \overline{m}_Q=m_Q^S(t^*) \times \overline{C}_m(t^*)\,. \label{matchi}\eeq
Equation (\ref{matchi}) holds also for $t^* \to \infty$, provided
at the same time $\delta \overline{m}(t^*) \to \delta  \overline{m}$.

At this order in $\alpha_s$
we can write
\beq \overline{m}_Q=\Bigl(M_H-{\cal E}+
\alpha_s(a)\frac{X}{a}\Bigr)
 \Bigl(
1-\frac{4\alpha_s(\overline {m}_Q)}{3 \pi} \Bigr)\label{matchid} \eeq
which is equivalent to the procedure where
the linearly divergent term is subtracted in perturbation theory.
Now both factors in eq.~(\ref{matchid}) contain renormalon singularities.
We can also rewrite eq. (\ref{matchid}) in the form
corresponding to the procedure where we use the unsubtracted,
linearly divergent ``pole" mass $M_H-{\cal E}$,
\beq \overline{m}_Q=\Bigl(M_H-{\cal E}\Bigr)
\Bigl(1-\frac{4\alpha_s(\overline {m}_Q)}{3 \pi}
+ \alpha_s(a)\frac{X}{a\overline{m}_Q}\Bigr)\ ,\label{matchid2} \eeq
where the divergent dependence on $a$ in the pole mass is compensated
by that in the  coefficient function. In this case no renormalon
singularities arise in higher orders, but the unsubtracted pole mass
and the coefficient function both contain power divergences. As
required, the relation (\ref{matchid2}) is independent of the
subtraction constant $ \delta \overline{m}(t^*)$.

\begin{figure}
\vspace{9pt}
\begin{center}\setlength{\unitlength}{1mm}
\begin{picture}(160,100)
\put(0,-55){\special{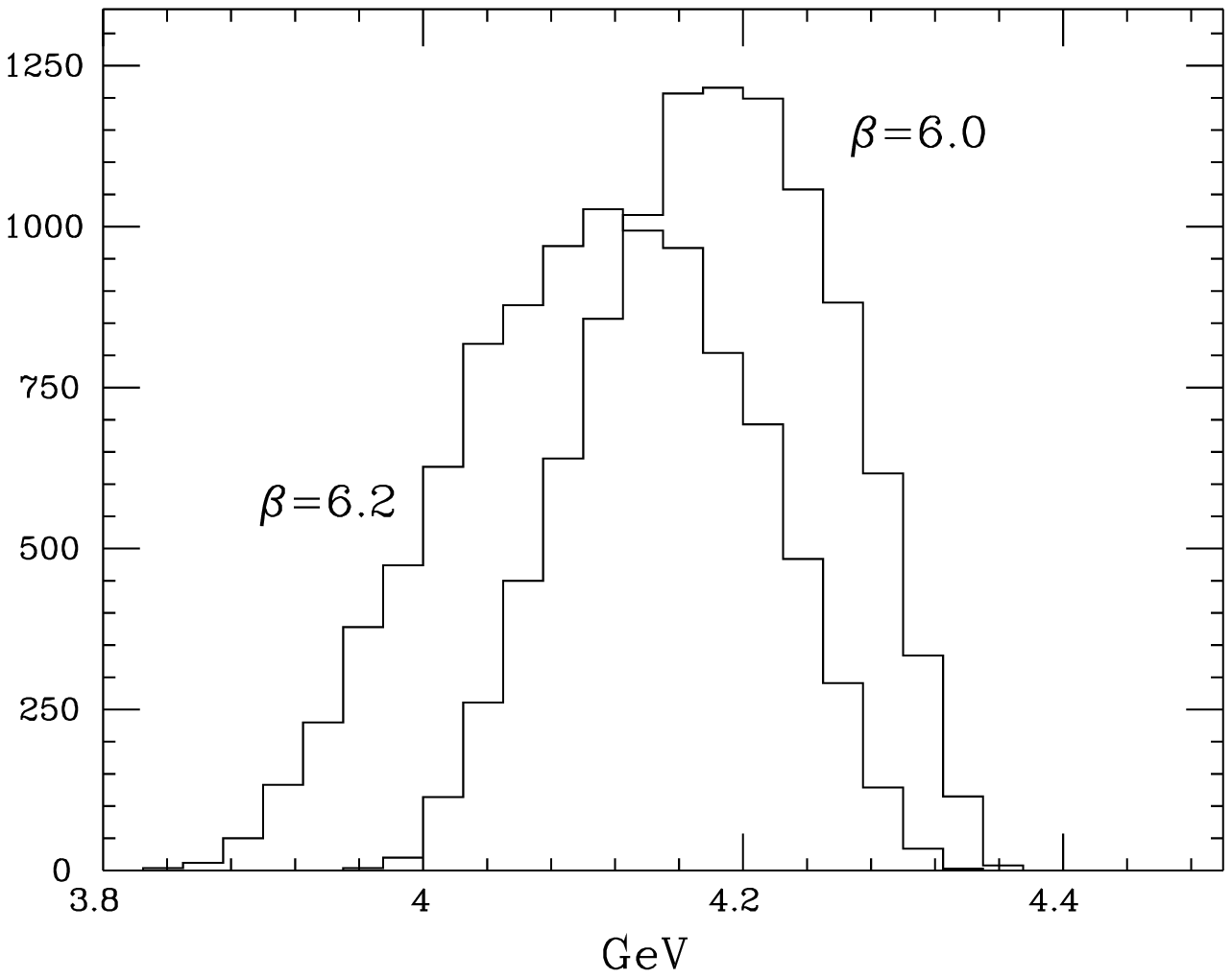}}
\end{picture}
\end{center}
\caption{\it{The ditribution of values of the $\overline{MS}$ $b$-quark mass
$\overline{m}_b$, for $t^* \to \infty$, at $\beta=6.0$ and
$\beta=6.2$. Similar distributions are obtained
for $\overline{m}_b$ using $t^*=3$--$7$.}}
\label{fig:hist}
\end{figure}

In the numerical evaluation of $\overline{m}_Q$ from
eqs. (\ref{matchi}) and (\ref{matchid}), we used
$M_B=5.278$; ${\cal E}$ from the
APE results, see subsection \ref{labarlead};
$a \, \delta \overline{m}(t^*)$ from eqs. (\ref{dbm60}) and (\ref{dbm62})
and table \ref{tab:lt};
$\alpha_s(a)$ was taken in the range  $0.13$ and $0.18$.
To obtain a distribution of values, we varied ${\cal E}$,
$\Lambda_{{\rm QCD}}$ and $a \, \delta \overline{m}(t^*)$
  according to a gaussian distribution; $a^{-1}$ was varied with
flat distribution within its error, while $\alpha_s(a)$ was written
in terms of the leading quenched expression of the running
coupling constant, evaluated at
the scale $\pi/a$, with $\Lambda_{{\rm QCD}}$ distributed according to
a flat distribution of width $\sigma$
 and such that $\alpha_s=0.13$ for $\Lambda_{{\rm QCD}}-\sigma$
and $\alpha_s=0.18$ for $\Lambda_{{\rm QCD}}+\sigma$.
The resulting distribution is a pseudo-gaussian, as can been seen
from fig. \ref{fig:hist} where two histograms of values of
$\overline {m}_b$, corresponding at $\beta=6.0$ and $6.2$, are shown.
{}From the width
of the distribution we estimate the average value and  error on
$\overline {m}_b$. Using eq. (\ref{matchi}),  we obtain at
$\beta=6.0$,
\beq
\overline {m}_b = 4.18 \pm 0.07 \hbox{ GeV  for } t^* \to \infty
\label{mqmsms1}\eeq
and
\beq
\overline {m}_b = 4.21 \pm 0.07 \hbox{ GeV  for } t^*=3-6 \ .
\label{mqmsms2}\eeq
The corresponding numbers at $\beta=6.2$ are
\beq
\overline {m}_b=4.11 \pm 0.09
 \hbox{ GeV  for } t^* \to \infty
\label{mqmsms3}\eeq
and
\beq
\overline {m}_b=4.13 \pm 0.09 \hbox{ GeV  for } t^*=4-7\ .
\label{mqmsms4} \eeq
Using eq. (\ref{matchid}),  we obtain instead
\beqn \overline {m}_b&=&4.22 \pm 0.07 \hbox{ GeV  at } \beta=6.0
\, ,\nonumber \\
\overline {m}_b&=&4.15 \pm 0.08 \hbox{ GeV  at } \beta=6.2
\label{mun} \eeqn
The difference between the results of eqs.
(\ref{mqmsms1})--(\ref{mqmsms4}) and
(\ref{mun}) can be interpreted as due to  higher order corrections
in $\alpha_s$. By combining the above results together we estimate
\beqn \overline {m}_b=4.17 \pm 0.05\pm 0.03 \hbox{ GeV}\,  \eeqn
where the second error is the systematic error coming from the different
methods used to extract $\overline {m}_b$ at this order in $\alpha_s$.
\subsection{Determination of $c_{1}$ and $c_{2}$} \label{sec:c12}
We have computed the ratio $\rho_{\vec{D}^{2}}(t)$, defined in eq.
(\ref{eq:c12}), using unsubtracted heavy-quark propagators, as
explained in section 2.2.  In order to do this calculation, we need the
expression of the heavy-quark propagator with the insertion of $\bar h
\vec D^2 h$
\be  \label{hevprobis}
\begin{array}{r@{\, =\,}ccl}
S^{a}(x \vert y)&\mbox{} {\displaystyle
\sum^{x^{4}}_{w^4=y^{4}}}\,
S\left(x,\vert\, w\right)\, \vec{D}^{2}_w(w^{4})\,
 S\left(w\,\vert \, y \right)\, ,
\end{array}
\ee
where the lattice heavy quark propagator $S\left(x \vert\, w \right)$
has been defined in eq. (\ref{ls0}), and in the improved case, we have
used the definition of the P-line given in  eq. (\ref{ipl}). For the
discretised version of $\vec{D}^{2}$ we have taken
\begin{eqnarray}
\left[\vec{D}_x^{2}\right]_{\alpha \beta} &=& \frac{1}{a^{2}}\,
\sum^{3}_{k=1}\, \left(\, U^{k}_{\alpha \beta}(x)\,
\delta_{x, x\, +\, a\,\hat{k}}
\,+\, U^{k\, \dag}_{\alpha \beta}(x\, -\, a\, \hat{k})\,
 \delta_{x, x\,
 -\, a\,\hat{k} }\,-\, 2\, \delta_{\alpha \beta} \, \delta_{x, x}\right) ,
\end{eqnarray}
\begin{table}
\centering
\begin{tabular}{||c|c|c|c|c||}
\hline
\hline
$\beta$ & time interval &$c_{1}$ & $c_{2}$&$\chi^2$/dof \\
\hline \hline
$6.0$& $4$--$12$& $0.06(2)$&-$0.759(6)$& $1.24$ \\
$6.0$& $5$--$12$& $0.01(5)$&-$0.748(10)$& $0.98$ \\
$6.0$& $5$--$14$& $0.01(5)$&-$0.748(10)$& $1.09$ \\
$6.0$& $6$--$14$& -$0.16(13)$&-$0.724(22)$& $0.80$ \\
 \hline  \hline
$6.2$& $4$--$12$& $0.10(4)$&-$0.698(9)$& $1.01$ \\
$6.2$& $5$--$12$& $0.16(9)$&-$0.708(18)$& $0.98$ \\
$6.2$& $5$--$14$& $0.17(9)$&-$0.710(17)$& $0.95$ \\
$6.2$& $6$--$14$& $0.37(18)$&-$0.739(29)$& $0.62$ \\
\hline  \hline
\end{tabular}
\caption{\it{ Results for the improved
renormalisation constants of the operator
$\bar h \vec D^2 h$ obtained by a linear fit to $\rho_{\vec{D}^{2}}(t)$.
The time interval of the fit is also given.}}
 \label{c2fit}
\end{table}

\begin{figure}
\vspace{9pt}
\begin{center}\setlength{\unitlength}{1mm}
\begin{picture}(160,100)
\put(0,-55){\special{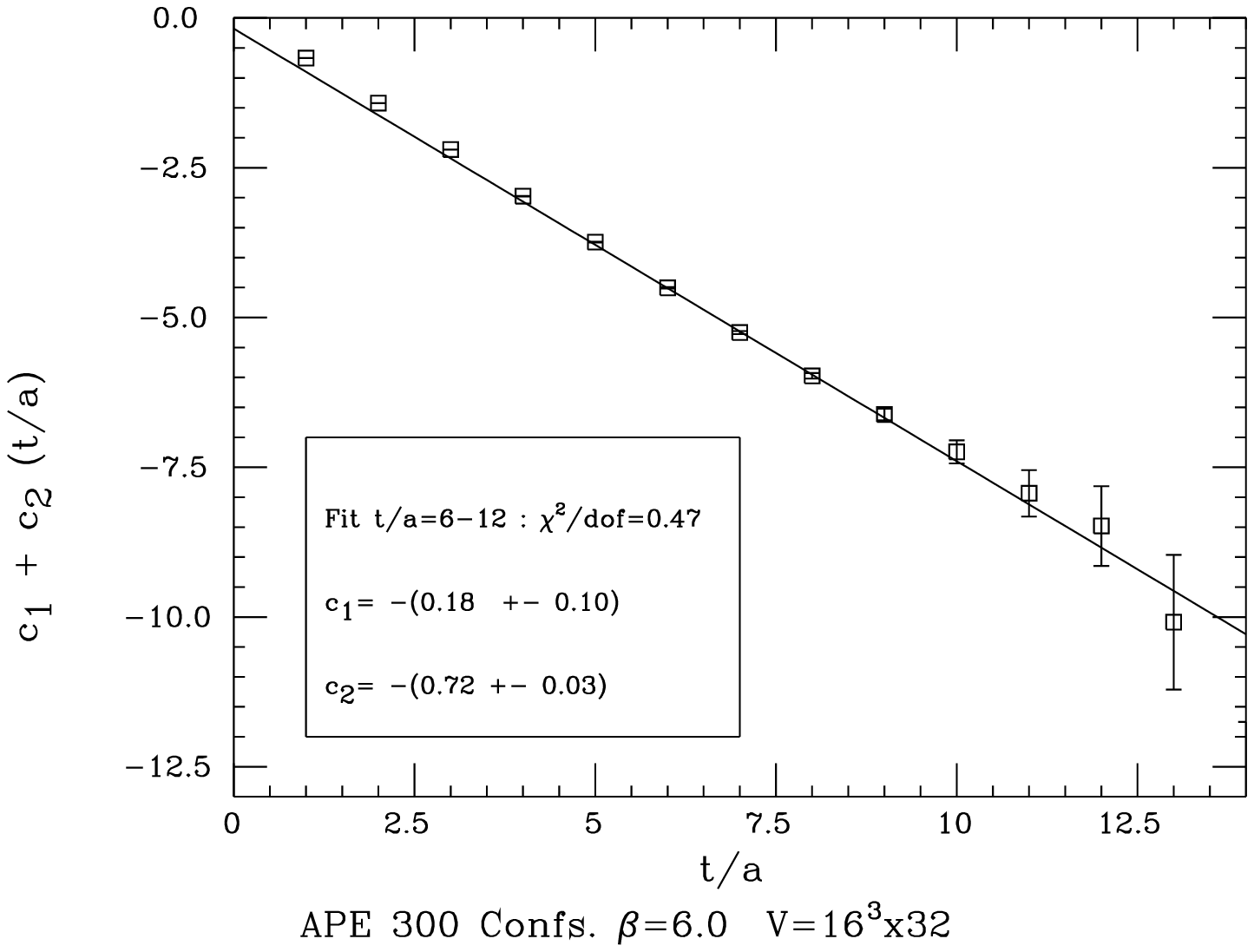}}
\end{picture}
\end{center}
\caption{\it{The ratio $\rho_{\vec{D}^{2}}(t)$ (improved case)
as a function of the time, at $\beta=6.0$ from \underline{set B}.
A linear fit of  $\rho_{\vec{D}^{2}}(t)$ to the expression
in eq. (\protect\ref{eq:twenty}) in the interval $6  \le t/a \le 12$,
is also given.}}
\label{fig:c1c2t}
\end{figure}
In fig. \ref{fig:c1c2t}, we plot $\rho_{\vec{D}^{2}}(t)$, as defined in
eq. (\ref{eq:c12}), as a function of the time $t$, at $\beta=6.0$  from
\underline{set B}.  In the same  figure,  we also give the result of a
linear fit  of  $\rho_{\vec{D}^{2}}(t)$ to eq. (\ref{eq:twenty}) in the
interval $6 \le t/a \le 12$. Similar results were obtained at
$\beta=6.2$ using the data  of \underline{set C}. The dependence of
$\rho_{\vec{D}^{2}}(t)$ on $t$ is in remarkable agreement with the
predicted linear behaviour~\footnote{ As expected, we found that the
results for $c_2$ in the improved and unimproved case are completely
compatible.  The latter are not reported here.}.  In order to monitor
the stability of the results, we have fitted $\rho_{\vec{D}^{2}}(t)$
using different time intervals  and in table \ref{c2fit} we show our
results for $c_1$  and $c_2$, in the improved case, at $\beta=6.0$ and
$6.2$. At $\beta=6.0$, we observe a shift of the value of $c_2$ towards
smaller values as we increase the  minimum  $t$-distance
($t_{{min}}=4,5,6$) at which the fit is performed. Since at
$\beta=6.2$, we find the opposite behaviour, i.e. the value of $c_2$ is
shifted towards larger values as  $t_{{min}}$ is increased, we believe
that the shift is a statistical effect rather than a systematic one.
{}From table \ref{c2fit}, we also observe that it is very difficult to
determine the value of $c_1$, which, for the improved propagator,
seems to be small, with a large relative error, and is very unstable
with respect to a change of the fitting interval. We expect that this
instability, which is correlated to the shift of the value of $c_2$ with
$t_{{min}}$, will be reduced with more accurate data for
$\rho_{\vec{D}^{2}}(t)$. Notice that $c_1$, unlike ${\cal E}$ and $c_2$
which are long-distance quantities, depends on the lattice
regularization, i.e. is different for the unimproved or the improved
heavy quark propagator. \begin{figure} \vspace{9pt}
\begin{center}\setlength{\unitlength}{1mm} \begin{picture}(160,100)
\put(0,-58){\special{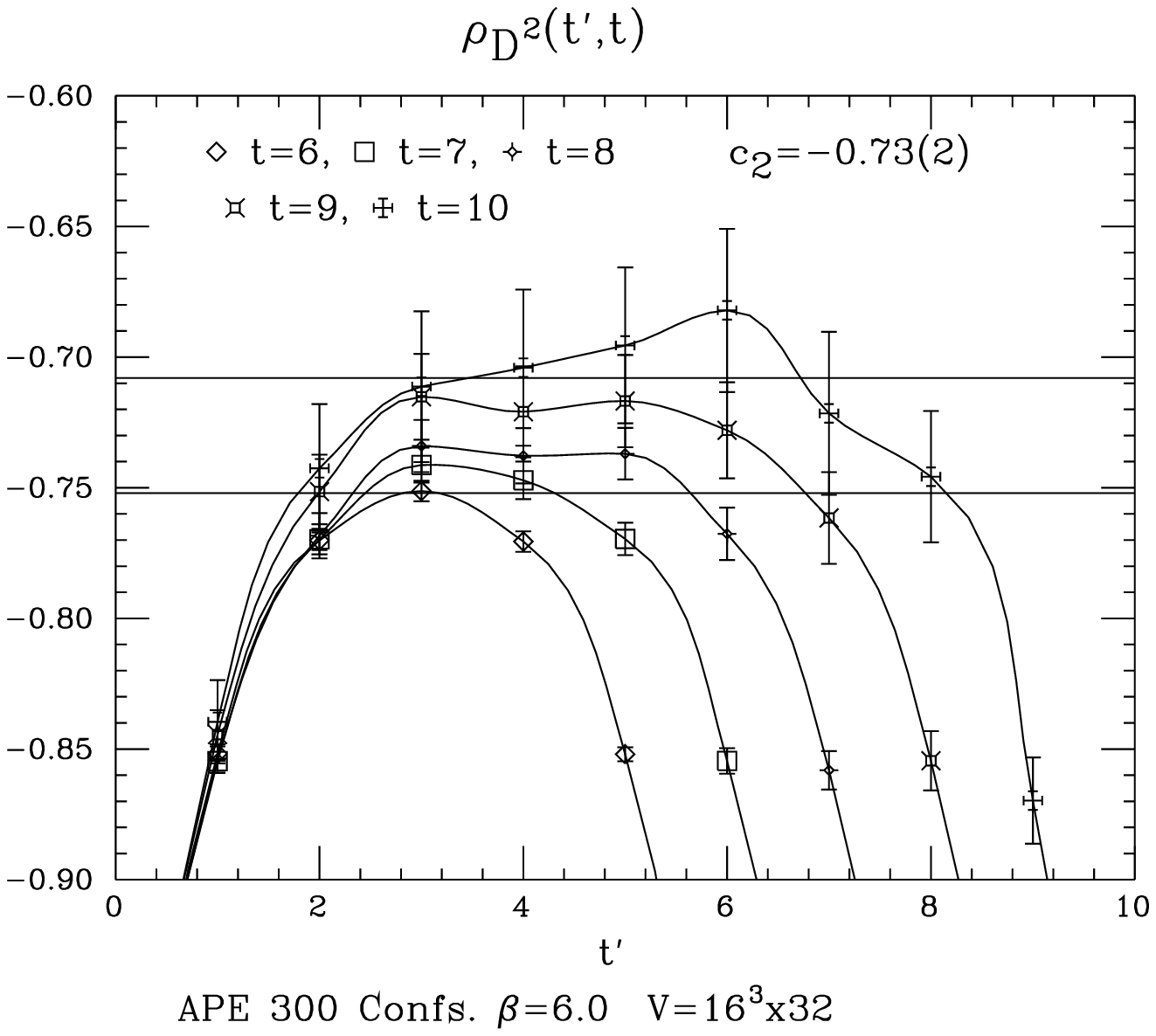}} \end{picture} \end{center}
\caption{\it{The ratio $\rho_{\vec{D}^{2}}(t^\prime,t)$ defined in eq.
 (\protect\ref{eq:c2tp}), in the
improved case, at $\beta=6.0$ from \underline{set B},
as a function of $t^\prime$,  at several values of $t$, $t=6$--$10$.}}
\label{fig:c2}
\end{figure}

We also present the results for $c_2=\rho_{\vec{D}^{2}}(t^\prime,t)$, at
$\beta=6.0$, obtained by  using  eq. (\ref{eq:c2tp}). In fig.
\ref{fig:c2},  we show $\rho_{\vec{D}^{2}}(t^\prime,t)$,  as a function
of $0 \le t^\prime \le t$, at several fixed values of $t$, $t=6$--$10$.
Up to contact terms, we expect $\rho_{\vec{D}^{2}}(t^\prime,t)$ to be a
constant in $t^\prime$, at fixed $t$, and also to be independent of
$t$. If the contact terms were entirely due to the mixing of the
kinetic energy operator with the inverse propagator,
eq.~(\ref{eq:d2ren}), we should find two spikes, at $t^\prime=0$  and
$t^\prime=t$, and a constant value of $\rho_{\vec{D}^{2}}(t^\prime,t)$
for $t^\prime \neq 0,t$. The presence of operators of higher dimension,
due to discretisation errors, introduces terms which behaves as
derivatives of $\delta$-functions (in time), giving rise to the
bell-shape behaviour of $\rho_{\vec{D}^{2}}(t^\prime,t)$ shown in fig.
\ref{fig:c2}. Thus in order to obtain $c_2$, we have to look for a
plateau in the  central region in $t^\prime$, at large values of $t$.
{}From the figure, we see that it is possible to recognize a plateau in
$t^\prime$ for $t=8$--$10$. At values of $t$ smaller than $t=8$, the
contact terms  are visible at all values of  $t^\prime$;  at  values of
$t$ larger than $t=10$ the statistical error become quite large. There
is a slight shift towards larger values of $c_2$ as $t$ is increased.
As discussed above, since the effect is opposite at $\beta=6.2$, we do
not believe that this is a systematic effect.
\begin{table}
\centering
\begin{tabular}{||c|c|c|c|c||}
\hline
\hline
$\beta$ & $t$& $t^\prime$ & $c_{2}$ &$\chi^2$/dof\\
\hline \hline
$6.0$& $6$ &$3$--$5$ &-$0.748(3)$&$17.00$ \\
$6.0$& $7$ &$3$--$5$ &-$0.735(6)$&$0.76$ \\
$6.0$& $8$ &$3$--$5$ &-$0.727(9)$&$0.14$ \\
$6.0$& $9$ &$3$--$6$ &-$0.713(15)$& $0.11$ \\
$6.0$& $10$ &$3$--$7$ &-$0.698(27)$& $0.16$ \\
 \hline \hline
$6.2$& $6$ &$3$--$5$ &-$0.674(4)$&$2.2$ \\
$6.2$& $7$ &$3$--$5$ &-$0.670(8)$&$0.03$ \\
$6.2$& $8$ &$3$--$5$ &-$0.680(12)$&$0.02$ \\
$6.2$& $9$ &$3$--$6$ &-$0.693(18)$& $0.40$ \\
$6.2$& $10$ &$3$--$7$ &-$0.723(29)$& $0.22$ \\
\hline \hline
\end{tabular}
\caption{\it{ Results for the
renormalisation constant $c_2$  computed from a weighted
average of  $\rho_{\vec{D}^{2}}(t^\prime,t)$ in $t^\prime$, at fixed $t$.}}
 \label{tab:c2tp}
\end{table}
In table \ref{tab:c2tp} we present the values of $c_2$, computed from a
weighted average of values of $\rho_{\vec{D}^{2}}(t^\prime,t)$ for
different $t^\prime$, at fixed $t$.
The average has been performed only in the central region, where there
appears to be a plateau. For the sake of comparison, we present the
results for several values of $t$, including small ones.

{}From the results given in  tables \ref{c2fit} and \ref{tab:c2tp},
and taking into account the previous discussion,
we believe that the best estimate of $c_2$ is obtained from
$\rho_{\vec{D}^{2}}(t^\prime,t)$, with $t=8$ and $t^\prime=3$--$5$
\be c_2 = -0.73 \pm 0.01 \pm 0.02 \hbox{ at }
 \beta=6.0 \, , \label{c260} \ee
\be c_2 = -0.68 \pm 0.01 \pm 0.02 \hbox{ at }
 \beta=6.2 \, , \label{c262} \ee
where the second error comes from the variation of the values of
$c_2$  with $t$.
These results can be compared with perturbation theory, which gives
$c_2\, =\,  - 5.19 \times  \alpha_s \sim -(0.67$--$0.93)$ for
$\alpha_s=-.13$--$0.18$.
For $c_1$ using the improved propagator, such a comparison is impossible,
due to the relatively large uncertainties in the non-perturbative
determination.

\subsection{Determination of the kinetic energy $\lambda_1$} \label{kinener}

The results for $\lambda_1$ have been obtained with limited statistics,
using the data of \underline{set A}. As explained in subsection
\ref{subl1}, the value of $\lambda_1^{{\rm bare}}$ can be obtained
from $R(t^\prime, t)$ as defined in eq. (\ref{eq:deltaasymp}). In
principle, we should evaluate  $R(t^\prime,t)$ using  the subtracted
propagators $ S^{ \prime}$. However, the argument used in section 2.2
for $\rho_{\vec{D}^{2}}(t)$ is also valid for $R(t^\prime,t)$, and
implies that we can obtain $R(t^\prime,t)$ by using the unsubtracted
heavy-quark propagators. In order to compute $R(t^\prime,t)$ we have
used single  and double cubic smeared interpolating operators $J=\bar h
\gamma_0 \gamma_5 q$, with smearing size  $L_s=7$, by using the heavy
and light quark propagators rotated into the Coulomb gauge. $L_s=7$ was
found to be the optimal value of $L_s$ for isolating the lightest meson
state at $\beta=6.0$ \cite{alltonw,allton}.

The procedure to extract operator matrix elements is standard. It is
the same as the second method that we used in the previous subsection
to determine $c_2$. At fixed $t$, we study the behaviour of the ratio
$R(t^\prime, t)$ as a function of $t^\prime$,  searching for a  plateau
in $t^\prime$. $\lambda_1^{{\rm bare}}$ is defined by  the  weighted
average of the data points in the central  plateau region, if this
exists. We will take as our best determination of  $\lambda_1^{{\rm
bare}}$,  the value  evaluated in a time interval where the ratio
$R(t^\prime, t)$ appears to be independent of both $t$ and $t^\prime$.
In addition, we have to require  that the  lightest state has been
isolated. With the smeared sources used in the present case, we know
that this happens at a time distance $(t-t^\prime)/a$ and $t^\prime /a
\ge 4-5$ from the source.  This implies that the total time distance
$t/a$ for $R(t^\prime, t)$ has to be at least $8$--$10$. Moreover,
using $(t-t^\prime)/a$ and $t^\prime /a \ge 4-5$, we eliminate the
contact terms, which on the basis of the discussion in the previous
subsection, cf. fig. \ref{fig:c2}, are expected to be present up to
distances of order $2$--$3$.
\begin{figure}
\vspace{9pt}
\begin{center}\setlength{\unitlength}{1mm}
\begin{picture}(160,100)
\put(0,-55){\special{PLAT3660.ps}}
\end{picture}
\end{center}
\caption{\it{The ratio $R(t^\prime,t)$ at $t/a=8$ as a function of the time
$t^\prime$. We show the value of $\lambda_1^{{\rm bare}}=\sum_{t^\prime/a
=3,5} R(t^\prime,t)$ (full orizontal line) and the relative band of error
(dashed orizontal lines).}}
\label{fig:d2}
\end{figure}
As an example of our results, we show in fig. \ref{fig:d2} the ratio
$R(t^\prime,t)$, at $t/a=8$, as a function of  $t^\prime$, the time at
which the  kinetic operator is inserted. With the present statistical
errors, it is not easy to identify the plateau region\footnote{ At
larger time distances, $t/a \ge 10$ the errors are even larger.}. If we
assume that we can use the central points ($t^\prime=3,4,5$) to extract
the value of the matrix element, we obtain for  the unrenormalised
value  $a^2 \, \lambda_1^{{\rm bare}}=-0.72 \pm 0.14$. This implies
that there is a large numerical  cancellation in the subtracted kinetic
energy, $ a^2 \,  \lambda_1=a^2 \,  \lambda_1^{{\rm bare}} - c_{2}=0.1
\pm 0.14$,  cf. eq. (\ref{eq:nextor}).  Due to the large statistical
and systematic errors and to the difficulty in the clear identification
of the plateau, it is not possible to obtain a value for the
renormalised kinetic energy from this simulation. We can only impose
the loose upper bound
$\lambda_1<1$ GeV$^2$ . Notice that in order to
reduce the statistical error to $0.1$ GeV$^{2}$, we need a sample about
$50$--$100$ times larger than our current one, corresponding to
$1500$--$3000$ gluon configurations.
Moreover, we would eventually also like to be able to extrapolate the
results to the chiral limit.
For
these reason, we are implementing the method described in this paper on
the 24 Gigaflops APE100 computer.

One could argue that the subtraction is not really necessary, since the
effective theory on the lattice does not have renormalons.  Even though
this is indeed true, the difficulty in the determination of
corrections  of order $1/m_Q$ related to the kinetic energy operator
would remain the same. The argument goes as follows. The bare kinetic
energy operator has a very large matrix element  $a^{-2} \times (a^2
\lambda_1^{{\rm bare}}) \sim 2^2 \times (-0.72)$ GeV$^2=-2.88$ GeV$^2$,
while one expects a correction due to the  kinetic energy of the heavy
quark of the order of the squared Fermi momentum $p_F^2 \sim
\Lambda_{{\rm QCD}}^2 \sim 0.1$ GeV$^2$. Thus the huge contribution of
the matrix element of the bare operator has to be compensated by the
corresponding term in the coefficient function of $\bar h h$. This
require an extreme accuracy in the  perturbative calculation of the
coefficient  function. This remains true in the subtracted as well as
in the unsubtracted case.

\section{Conclusions} \label{conclu}

In this study we have shown that the method for the non-perturbative
renormalisation of the lattice operators $\energy$ and $\kkinetic$,
proposed in
ref.  \cite{d2reno}, is feasible in current computer
simulations. We have been able to obtain the
subtraction constants
of the operators $\kkinetic$ and $\energy$ with a small statistical
error
(in the former case, particularly for the constant $c_2$ which is
needed for many physical applications).
The binding energy of the B-meson, $\labar$, has been also
calculated with an error of about
$15 \%$ and was found to be significantly smaller than other estimates,
based on different definitions \cite{fnal,shif}. We have also computed
the kinetic energy of the heavy quark in the B-meson.
With our current statistical sample we can only impose the bound on
$\lambda_1 < 1$\, GeV$^2$, on the matrix element of the kinetic energy
operator. This is due to the large numerical cancellation when the
counter-term is subtracted. We are planning to improve the precision
of our results
by using a much larger sample of gluon configurations, and hopefully
to obtain a significant result for $\lambda_1$.

Our preferred determination of $\lb$ was based on the behaviour of
the heavy quark propagator at large times. It is important to verify
the validity of the condition (\ref{eq:condition}) by extending the
calculation of $\delta\overline{m}(t)$ to larger values of $t$. This
requires a high-statistics simulation on a large lattice, and we are
currently undertaking such a study. The results will be reported
elsewhere.

The present study concerned some important matrix elements which appear
in the HQET. We were able to determine $\lb$, defined in different
prescriptions, with good precision. This  encourages us to extend the
calculation to  other matrix elements which appear at $O(1/m_Q)$, and
beyond, in the HQET.  The main limitation to the matching to the full
theory is due to the fact that the relevant Wilson coefficients have
only been computed at first order in $\alpha_s$. We are planning to
extend these calculations to higher orders. One may also extend the
present approach to  matrix elements which appear at higher orders in
other important operator expansions, such as the non-leading twist
operators in deep inelastic scattering or higher dimensional
condensates used in QCD sum-rules.

\section*{Acknowlegments}

We acknowledge the partial support by the EC contract
CHRX-CT92-0051.
M.C. and G.M. acknowledge the partial support by M.U.R.S.T.
V.G.~wishes to thank the Istituto di Fisica ''G. Marconi'' of the
Universit\`a di Roma ''La Sapienza'' for its hospitality and
acknowledges the European Union for their support through the award of a
Postdoctoral Fellowship (EC contract CHRX-CT93-0132).
C.T.S. acknowledges the Particle Physics and Astronomy Research Council
for its support through the award of a Senior Fellowship.

\end{document}